# Light Induced Electron Spin Resonance Properties of van der Waals $CrX_3$ (X = Cl, I) Crystals

S. R. Singamaneni[1*], L. M. Martinez[1], J. Niklas[2], O. G. Poluektov[2], R. Yadav[4,5], M. Pizzochero[4,5], O. V. Yazyev[4,5], and M. A. McGuire[3]

[1]Department of Physics, The University of Texas at El Paso, El Paso, Texas 79968, USA
[2]Chemical Sciences and Engineering Division, Argonne National Laboratory, Lemont, Illinois 60439, United States
[3]Materials Science and Technology Division, Oak Ridge National Laboratory, Oak Ridge, Tennessee 37831, USA
[4]Institute of Physics, Ecole Polytechnique Fédérale de Lausanne (EPFL), CH-1015 Lausanne, Switzerland
[5]National Centre for Computational Design and Discovery of Novel Materials (MARVEL), Ecole Polytechnique Fédérale de Lausanne (EPFL), CH-1015 Lausanne, Switzerland

The research on layered van der Waals (vdW) magnets is rapidly progressing owing to exciting fundamental science and potential applications. In bulk crystal form, $CrCl_3$ is a vdW antiferromagnet with in-plane ferromagnetic ordering below 17 K, and $CrI_3$ is a vdW ferromagnet below 61 K. Here, we report on the electron spin resonance (ESR) properties of $CrCl_3$ and $CrI_3$ single crystals upon photo-excitation in the visible range. We noticed remarkable changes in the ESR spectra upon illumination. In the case of $CrCl_3$, at 10 K, the ESR signal is shifted from g = 1.492 (dark) to 1.661 (light), line width increased from 376 to 506 Oe, and the signal intensity is reduced by 1.5 times. Most interestingly, the observed change in the signal intensity is reversible when the light is cycled on/off. We observed almost no change in the ESR spectral parameters in the paramagnetic phase (>20 K) upon illumination. Upon photo-excitation of $CrI_3$, the ESR signal intensity is reduced by 1.9 times; the g-value increased from 1.956 to 1.990; the linewidth increased from 1170 to 1260 Oe at 60 K. These findings are discussed by taking into account the skin depth, the slow relaxation mechanism and the appearance of low-symmetry fields at the photo-generated $Cr^{2+}$ Jahn-Teller centers. Such an increase in the g-value as a result of photo-generated $Cr^{2+}$ ions is further supported by our many-body wavefunction calculations. This work has the potential to extend to monolayer vdWs magnets by combining ESR spectroscopy with optical excitation and detection.

*srao@utep.edu

The van der Waals (vdW) layered magnets such as $CrX_3$ (X = Cl, I) with a unique atomic-level exfoliable structure and rich physical properties (e.g. ferromagnetism at the atomic limit) are promising materials for next generation spintronic and magneto-electronic applications[1–5]. Their outstanding features such as the interplay of dimensionality, correlation, charge, orbital character, and topology of being susceptible to a large variety of external stimuli makes the versatile control of 2D magnetism possible by electrical, chemical, and optical approaches[4]. Electric field, electrostatic doping, and mixed halide chemistry have been shown to control the magnetic properties (magnetization, coercive field, and magnetic order) of these materials[2,3,6]. Unlike the above approaches, light is a particularly intriguing tool, which could potentially enable remote and rapid control of magnetic properties of layered vdW magnets which are susceptible to external stimuli, with much less energy consumption to develop spin electronics.

To advance the frontier of knowledge, photo-excitation investigations need to be extended to the above vdW materials. Numerous theoretical works[7-10] predicted that there is a strong coupling between light and magnetism and strong magneto-optical effects[7] in vdW magnets due to complete spin polarization in conduction and valence bands. Previous researchers have measured the magnetic properties of these layered materials using light[1–5]. However, controlling their magnetic properties using light has remained largely unexplored. Since the magnetic properties of bulk crystals serve as a basis for the understanding of magnetic phenomenon in reduced dimensions, e.g., in few or monolayer $CrX_3$, a profound knowledge and understanding is evidently needed, especially considering that magnetic mono and bilayers of $CrX_3$ have now become accessible.

In $CrX_3$, $Cr^{3+}$ ions are arranged in a honeycomb network and located at the centers of edge sharing octahedral of six halogen atoms. $CrI_3$ is a ferromagnet[11] with a Curie temperature ($T_C$) =

61 K, and CrCl$_3$ is an in-plane ferromagnet[12] and out-of-plane antiferromagnet (AFM) with an ordering temperature (Néel temperature, T$_N$) near 17 K. To note, with the application of few hundred Gauss magnetic field, CrCl$_3$ can be turned into a ferromagnet[13]. Therefore, most likely, CrCl$_3$ turns to a complete ferromagnet during the X-band ESR measurements (not before the magnetic field was swept) as several thousand Gauss magnetic fields are applied to detect the Electron spin resonance (ESR) signal. In essence, during the ESR measurements, the ground states of these two compounds can be approximated as ferromagnetic state.

ESR spectroscopy is an indispensable technique for studying magnetic interactions in materials such as CrX$_3$ containing unpaired electron spins. ESR spectroscopy has been extensively employed to study the spin interactions in other low dimensional materials such as graphene nanoribbons, carbon nanotubes, and other vdW crystals[14-20]. There have been no prior reports appearing in the literature on the study of the photo-excited ESR properties of CrX$_3$ though they are known to be magnetically and optically active. In this letter, by employing experimental and theoretical approaches, we report that the ESR properties –reflective of local magnetic exchange interactions in CrX$_3$ can be tuned through optical means.

The experimental materials, methods, and procedures for quantum chemistry calculations are discussed in the Supplementary Materials (SM). To trace the ferromagnetic ordering temperature (T$_C$), the temperature dependent magnetic measurements were performed (under dark) on CrCl$_3$ single crystals, and confirmed that the magnetic order appeared near 17 K as shown in Fig. 1(a), which is consistent with the literature[12]. Isothermal magnetization measurements (inset of Fig. 1(a)) show the expected soft ferromagnetic nature of this compound. As shown in SM, Figs. 1 (a) and (d) present the X-band (9.45 GHz) ESR spectra (0-7 kOe) recorded on CrCl$_3$ at 10 K and 20 K when the light is ON (shown in red) and OFF (shown in black) in the ferromagnetic phase, respectively. Also included is the light on minus light off signal (shown in blue). In SM, Figs. 1 (b) and (c) depict the ESR signal intensity as a function of time (in seconds) when the light is ON, OFF, and ON again. As it can be noted, at 10 K, the signal intensity at 3870 Oe drops to a much lower stationary value in less than half a second once the light is OFF and the signal is completely reversible to its original level once the light is ON again in few milliseconds which is remarkably close to the (radiative luminescence) lifetime (13-ms) reported in the literature[21,22] for $Cr^{3+}$. The stationary value reached during the photo-excitation was found to depend on the intensity of the light. That means light is strongly coupled with the magnetic behavior in $CrCl_3$.

The ESR spectral parameters such as g-value, line width, and integrated intensity for both light ON and OFF signals, as plotted in Figs. 1(b-d) as a function of temperature, were obtained from the computer-generated fits using a Dysonian line shape (SM Fig. 2). As the data suggest, distinct ESR spectral parameters are noted upon photo-excitation in the ordered magnetic phase (<17 K) of $CrCl_3$. The spectral parameters are the same in the paramagnetic phase (>17 K) of $CrCl_3$. Several interesting features can be noted. Firstly, the signal is strongly shifted from g = 1.492 (dark) to 1.661 (light). Secondly, the ESR signal is broadened from 376 Oe to 506 Oe. Thirdly, the ESR signal intensity is reduced by 1.5 times. The obtained ESR spectral parameters (resonance field/g-value, linewidth) under dark are the benchmark signatures of $Cr^{3+}$ (S = 3/2) ions in the octahedral site, consistent with the previous reports[23] on this compound, and not related to any defect related spin centers. ESR signal intensity is decreased by increasing the incident light intensity that can be attributed to the change in the skin depth[24]. The photo-magnetic effects appeared only below the ordering temperature of 17 K (SM Figure 1) and increased with a decrease in the temperature.

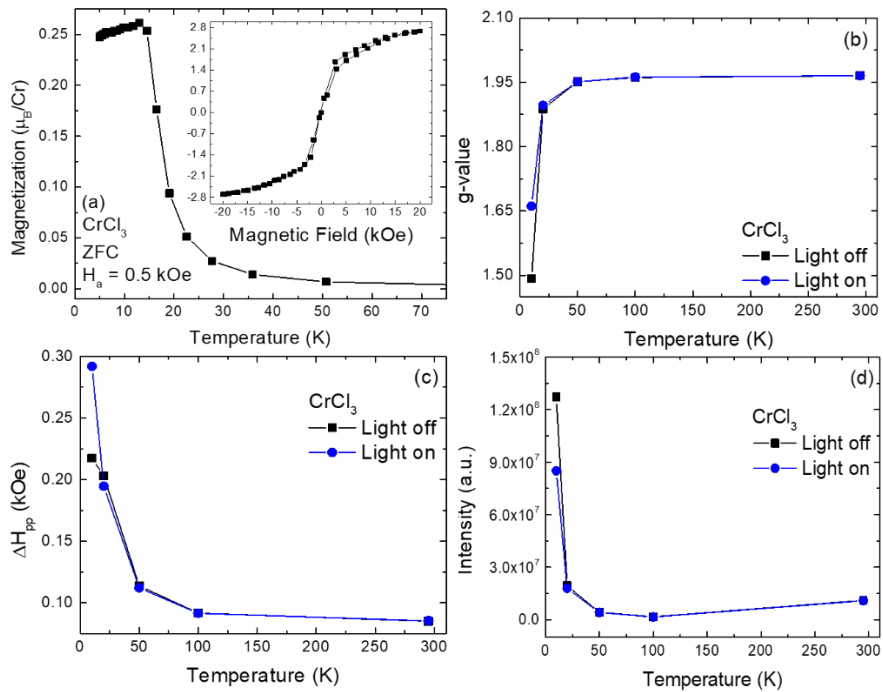

**Figure.1**: (a) The Zero-Field-Cooled (ZFC) temperature dependence of the magnetization collected from $CrCl_3$ in the out-of-plane direction with an applied field ($H_a$) of 0.5 kOe. The magnetic order appeared at 17 K, consistent with Ref. 12 . The inset shows the isothermal magnetization curve of $CrCl_3$ measured at 2 K. The temperature dependences of g-value (b), the intensity (c), and the linewidth (d) of the ESR signal collected with light ON (in blue) and OFF (in black) from $CrCl_3$.

The ESR signal is also measured at and above 20 K in the paramagnetic phase, and the data are plotted in SM Figs. 1 (e-g) at the temperatures of 50 K, 100 K, and 295 K, respectively. As it can be noted, the ESR signals before and after the light is ON almost overlap. It should be mentioned that the pure heating effect takes at least several seconds, however, the response in the present sample is much faster (<0.5 seconds). Heating mainly comes from IR light, but our LED emission does not contain an IR component. Weak signals appeared at the low field side of the spectrum collected at 10 K, which could be due to the small ferromagnetic inhomogeneity or isolated $Cr^{3+}$ present[25] in the crystal and disappeared as the measurement temperature is increased.

Now, we will present our data on CrI$_3$. To determine T$_C$, the magnetization (dark) was recorded as a function of temperature and magnetic field. As plotted in Fig. 2(a), CrI$_3$ shows a clear T$_C$ at 60 K. Also, the magnetic hysteresis loop changes its curvature from ferromagnetic to paramagnetic across the magnetic phase transition at 60 K (inset of Fig. 2(a)), consistent with previous works[11].

Once we knew the T$_C$, the light induced ESR measurements were performed on CrI$_3$ across T$_C$ (10 K, 30 K, 60 K, 65 K, and 100 K). Similar to the case of CrCl$_3$, the light in the visible region is employed for excitation. First, we will begin discussing the data collected in the paramagnetic phase (>60 K). Quite strikingly, unlike in the case of CrCl$_3$, we observed significant changes in the ESR spectral properties in the paramagnetic phase. The data are plotted in SM Figs. 3 (a-e). The ESR signals were fitted with the Dysonian line shape equation[26] (SM Fig. 4) and the ESR spectral parameters are plotted in Figs. 2 (b-d) as a function of temperature. The ESR signal is shifted from g = 1.956 to 1.990 (Fig. 2(b)) and the signal is broadened from peak-to-peak line width of 1170 to 1260 Oe (Fig. 2 (c)). Most notably, upon photo-excitation, we noticed that there is a stronger quenching of ESR signal intensity by 1.9 times measured at 60 K (Fig 2(d)). As presented in SM, Figs. 3 (d) and (e) show the time dependence of ESR signal intensity when the light is ON and OFF measured at 60 K and 100 K, respectively. The decrease in ESR signal intensity is directly proportional to the intensity of incident light. Most importantly, the ESR signal is reversible immediately upon switching OFF the light. This observation is consistently noticed even at other higher temperatures of 65 and 100 K in the paramagnetic phase. However, in the ferromagnetic phase (< 60 K), we found that ESR signals are irreproducible (SM Fig. 5). Most likely, this behavior could be due to the strong interaction of light with magnetic domains which causes instability of ferromagnetic domains in CrI$_3$[27]. The measurement was repeated even after cooling the sample under the magnetic field of 3000 Oe (to produce single ferromagnetic domain).

However, we noticed that the results remain unaffected. Upon closer inspection, the g-value and linewidth obtained from these two compounds are quite different and is attributed to magnetic anisotropy arising mostly from the spin-orbit coupling on $I^-$ which exceeds that on $Cr^{3+}$ or $Cl^-$ by more than an order of magnitude[11-13,28].

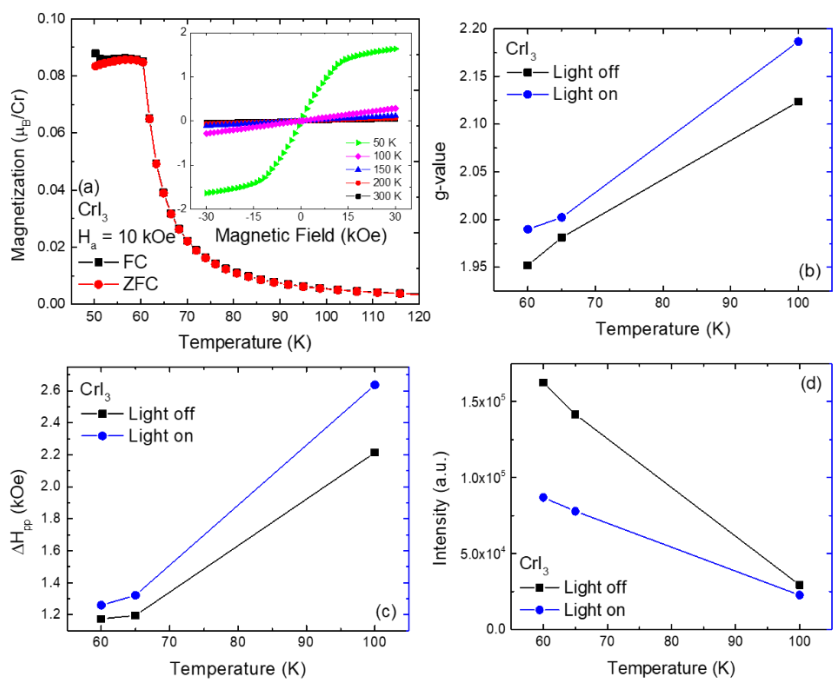

Figure 2: (a) Variation of (in-plane) magnetization as a function of magnetic field (M vs. H) recorded at various temperatures on $CrI_3$. Inset shows the ZFC and Field-Cooled (FC) temperature dependent magnetization curves collected with an applied field ($H_a$) of 10 kOe, consistent with Ref. 11. The temperature dependences of g-value (b), the linewidth (d), and the intensity (c) of the ESR signal collected with light ON (in blue) and OFF (in black) from $CrI_3$.

We have studied the magnetic properties of $CrI_3$ upon photo-excitation in ferromagnetic (50 K, 3T – instrument limit) and paramagnetic regions (100 K) as plotted in Fig. 3. The data are plotted and compared with that of pristine (without light) $CrI_3$. The isothermal (50 K) magnetization of pristine $CrI_3$ is consistent (1.8 $\mu_B$/Cr, slightly lower than 2 $\mu_B$/Cr) with the previous work[11]. The results show that the magnetization (at 30 kOe) increased upon photo-excitation with ligand to metal charge transitions LMCT 1e (642 nm) and LMCT 2e (459 nm). It is most likely that the increase in magnetization (at 50 K) corresponds to the (in part) formation of $Cr^{2+}$ (S = 2). The increase in the magnetization upon 642 nm photo-excitation is also reflected in the increase (3.9 to 4.9 $\mu_B$/Cr) of the effective magnetic moment ($\mu_{eff}$) obtained through Curie-Weiss fits. Therefore, as discussed later in the manuscript, the magnetic behavior can be due to the combination of $Cr^{3+}$ and $Cr^{2+}$ and form two sub-magnetic systems.

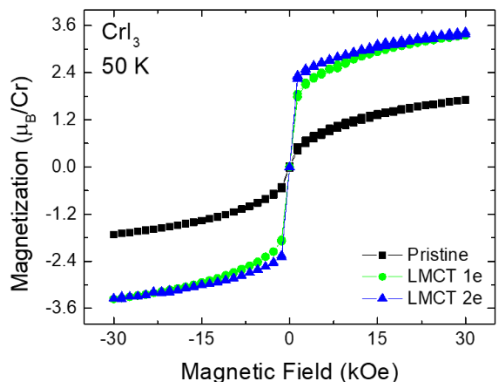

**Figure 3:** Isothermal (50 K, 3T) magnetization collected on $CrI_3$ upon photo-excitation with ligand to metal charge transition LMCT 1e (642 nm) and LMCT 2e (459 nm). For comparison, the data obtained from the dark state (pristine) is also included.

Based on our experimental findings, we attempt to understand the effect of light on the ESR spectral properties of $CrX_3$. We will first consider the case of $CrCl_3$. As can be seen, almost no changes in the paramagnetic phase of this material upon illumination were observed, and hence we will not discuss this further. Instead, we will focus our efforts in its magnetically ordered phase.

Our experimental findings presented here resemble those of previous works[24,29,30], where the photo-magnetic effects were reported on chromium-based chalcogenide ferromagnetic semiconductor, namely, $CdCr_2Se_4$ associated with $Cr^{3+}$. The authors found that the ESR spectral properties such as ESR signal intensity, resonance field and signal width are modified upon photo-excitation in the ferromagnetic phase (<130 K) of $CdCr_2Se_4$; and no changes were observed in the paramagnetic phase (>130 K). This situation is exactly similar to the present case. That leads us to adopt the mechanisms here.

The photo-induced electronic transitions create $Cr^{2+}$ ions in $CrX_3$ according to the scheme $Cr^{3+}$ + $e^-$ → $Cr^{2+}$. The photo-activated electrons from the valence band are trapped on the octahedral $Cr^{3+}$ sites creating $Cr^{2+}$ centers. They are assumed to be formed between the valence band and $Cr^{2+}$ energy level situated below the bottom of the conduction band.

The experimental results can be interpreted within the framework of a band model[24] (SM Fig. 6). At a distance of 3.02 eV from the valence band, there is a narrow conduction band consisting of $Cr^{2+}$ ($3d^4$) electrons. The transition between these two bands correspond to a change in the chromium-ion valence according to the scheme $Cr^{3+}$ + $e^-$ → $Cr^{2+}$. The observed light induced changes in the ESR parameters may be caused by photo-induced electron transitions between the valence band and localized $Cr^{2+}$ ($d^4$, S = 2) levels situated at a distance of 0.08 eV below the bottom of the conduction band. The observed light-induced decrease in ESR signal intensity may be caused by the photo-induced electron transitions between the above mentioned $Cr^{2+}$ levels and broad conduction band from which the recombination process occurs much faster. It should be noted that the $Cr^{2+}$ ESR signal could not be detected at X-band frequency of 9.45 GHz at any measured temperatures, due to the shorter relaxation time and high zero field splitting[31].

The changes in the g-value and signal width could be explained by taking into account the following two different mechanisms: (i) the slow relaxation mechanism in which the observed changes are attributed to $Cr^{2+}$, and (ii) the appearance of random low-symmetry fields acting on the Jahn-Teller $Cr^{2+}$ ions on the octahedral sites. In order to assess whether the observed increase in the g-value is related to the photo-generated $Cr^{2+}$ ions, we performed quantum chemistry calculations on finite-size models derived from the experimental crystal structure[11,12]. The models consist of a Cr-centered octahedral unit treated at the correlated level surrounded by the three nearest-neighbor octahedral units treated at the Hartree-Fock level. Such models are further embedded in an array of point charged fitted to the Madelung potential of the corresponding crystal lattice. As a first step, we have obtained multiconfiguration wavefunctions through complete-active-space self-consistent-field (CASSCF) computations[32]. This was accomplished using an active space consisting of either three (for $Cr^{2+}$ ions) or four (for $Cr^{3+}$ ions) electrons residing in the three $t_{2g}$ and two $e_g$ orbitals at the central Cr site. CASSCF wavefunctions were optimized for the lowest quartet and five doublet states (one triplet, two quintet and three singlet states) in the case of $Cr^{3+}$ ($Cr^{2+}$) ions, which subsequently entered the spin-orbit treatment to yield spin-orbit coupled states. As a second step, single- and double-excitations from the Cr 3$d$ and ligands $p$ valence shells are accounted for in the multireference configuration-interaction (MRCI) calculations[33,34]. g-values were obtained by means of the methodology devised[35] for the ground state (S = 3/2) of both $CrCl_3$ and $CrI_3$ systems, in which the Cr ion features a formal 3+ oxidation state. We remark that a similar computational strategy was successfully adopted in earlier quantum chemistry studies for a number of honeycomb lattice systems[36–39], including monolayer $CrI_3$[39].

We found a g-value of 1.42 for $CrCl_3$ and 1.92 for $CrI_3$. In addition, we performed calculations assuming $Cr^{2+}$ octahedral units, obtained by adding an extra electron at the Cr site and properly adapting the embedding to ensure charge neutrality. The g-values obtained in the case of $Cr^{2+}$ ions are 1.79 for $CrCl_3$ and 2.08 for $CrI_3$. This drastic change in the g-values is in line with the experimental findings discussed above, hence indicating that the reduction of $Cr^{3+}$ to $Cr^{2+}$ ions is at the origin of the observed photo-induced evolution of the ESR spectra. This is also supported by our recent ultrafast optical pump-probe measurements (not shown) on $CrI_3$. We suggest that the slight discrepancy between the quantum chemistry and experimental data can be traced back to the fact that only a fraction of the $Cr^{3+}$ ions are converted into the $Cr^{2+}$ ions upon photo-excitation.

As learned from the above discussion, the system under investigation consists of two subsystems. The first is the ferromagnetically ordered $Cr^{3+}$ ions, in which, the spin-orbit coupling is known to be quenched. The second subsystem is formed by photo-generated $Cr^{2+}$. This ion is characterized by strong spin-orbit coupling (40-60 cm$^{-1}$). These two subsystems are coupled by an exchange interaction that, in general, is anisotropic. The interaction between these two subsystems can slow down the spin dynamics and cause the increase in linewidth as well as g-value due to the introduction of spin-orbit coupling from Jahn-Teller distorted $Cr^{2+}$ ion.

Similar mechanisms can be extended to the case of photo-induced ESR spectral properties of $CrI_3$, in which, the light induced changes are much stronger. In addition, the g-value and the linewidth in $CrI_3$ are much higher than those of $CrCl_3$ due to strong spin-orbit coupling of I as well as ligand induced giant magnetic anisotropy[13]. The photo-induced ESR spectral properties in the ferromagnetic phase of $CrI_3$ could not be reproduced. It can happen that the light induced $Cr^{2+}$ centers strongly interact with the ferromagnetic domains and modify the domain wall thickness and domain-wall mobility. That can cause the instability in the domain structure as was observed[27] in the case of $FeBO_3$. However, the question as to why the photo-induced ESR spectral properties are not observed in the paramagnetic phase of $CrCl_3$ remains to be answered.

To conclude, we reported remarkable changes in the ESR spectra from $CrX_3$ upon illumination. In the case of $CrCl_3$, in the antiferromagnetic phase, with the light ON, it is observed that the ESR signal is strongly shifted from g = 1.492 (dark) to 1.661 (light), significantly broadened the signal from 376 (dark) to 506 Oe, and the signal intensity is strongly reduced by 1.5 times. Most importantly, the signal intensity is found to be completely reversible. In the case of $CrI_3$, drastic changes were noted upon photo-excitation with the visible light. The ESR signal intensity is reduced by 1.9 times; the g-value increased from 1.956 to 1.990; the linewidth increased from 1170 to 1260 Oe in the paramagnetic phase. Most likely, the photo quenching effect, the slow relaxation mechanism and the appearance of low-symmetry fields at the photo-generated $Cr^{2+}$ Jahn-Teller centers, supported by theoretical calculations, could explain the changes in ESR spectral properties upon photo-excitation. This effort forms a significant step forward toward extending this type of work to mono- and bilayers of $CrX_3$, which may provide unprecedented opportunities to study the light-induced magnetism in the two-dimensional limit.

Supplemental Materials

See supplementary materials for the experimental methods and materials, magnetic properties, additional ESR measurements and analysis, and quantum chemical calculations performed on $CrCl_3$ and $CrI_3$.

Acknowledgements

S.R.S. and L.M.M. acknowledge support from a UTEP start-up grant. L.M.M. acknowledges the NSF-LSAMP BD fellowship. Crystal growth and characterization at Oak Ridge National Laboratory (M.A.M.) was supported by the US Department of Energy, Office of Science, Basic Energy Sciences, Materials Sciences and Engineering Division. J.N. and O.G.P. gratefully acknowledge support by the U.S. Department of Energy, Office of Science, Office of Basic Energy Sciences, Division of Chemical Sciences, Geosciences, and Biosciences, through Argonne National Laboratory under Contract No. DE-AC02-06CH11357. R. Y. is supported by the Sinergia Network NanoSkyrmionics (Grant No. CRSII5-71003). M. P. is supported by the Swiss National Science Foundation (Grants No. 162612 and 172543). Calculations were performed at the Swiss National Supercomputing Centre (CSCS) under the project s832.

## Data Availability Statement

The data that supports the findings of this study are available within the article (and its supplementary material).

15**Supplementary Materials (SM):**

**Light Induced Electron Spin Resonance Properties of van der Waals CrX$_3$ (X = Cl, I) Crystals**

S. R. Singamaneni[1*], L. M. Martinez[1], J. Niklas[2], O. G. Poluektov[2], R. Yadav[4,5], M. Pizzochero[4,5], O. V. Yazyev[4,5], and M. A. McGuire[3]

[1]Department of Physics, The University of Texas at El Paso, El Paso, Texas 79968, USA
[2]Chemical Sciences and Engineering Division, Argonne National Laboratory, Lemont, Illinois 60439, United States
[3]Materials Science and Technology Division, Oak Ridge National Laboratory, Oak Ridge, Tennessee 37831, USA
[4]Institute of Physics, Ecole Polytechnique Fédérale de Lausanne (EPFL), CH-1015 Lausanne, Switzerland
[5]National Centre for Computational Design and Discovery of Novel Materials (MARVEL), Ecole Polytechnique Fédérale de Lausanne (EPFL), CH-1015 Lausanne, Switzerland
# Experimental Methods and Materials:

The crystals used in this study were grown by chemical vapor transport (CVT) method as described earlier[1,2] by one of us (M.M) at Oak Ridge National Laboratory. Thin, large violet-colored transparent plate-like single crystals of CrCl$_3$ with lateral dimensions up to several millimeters were grown by recrystallizing commercial CrCl$_3$ using CVT. A similar procedure was adopted to prepare CrI$_3$. For magnetization measurements, performed in the out of plane direction, we employed two Quantum Design Magnetometers: Physical Property Measurement System (PPMS) (2-400 K, ±70 kOe) and versa lab VSM magnetometer (50-400 K, ±30 kOe). For electron spin resonance (ESR) measurements, one platelet is taken with arbitrary direction, and the magnetic field was applied perpendicular to the plane of the plate within 10-degree accuracy. For ESR measurements, the modulation frequency was 100 kHz, and the modulation amplitude was 8-10 Oe. The microwave power was 8 µW (200 mW source) with 44 dB attenuation.

The photo physics of chromium (III) compounds is well known[3–6]. The optical absorbance of a CrCl$_3$ crystal showed the approximate color of the visible light for photon energies between 1.7 and 3.3 eV[2]. The results are consistent with the data reported in the thorough study of the optical properties of this material by Pollini and Spinolo[5]. A band gap of 3.1 eV is approximated by the onset of strong absorption at higher energies. Below the band edge, there are two broad absorptions in the red and green centered near 1.7 and 2.3 eV, to which the violet color of the crystal can be attributed. Hence, we choose to excite this material with visible light (1.8 to 3.1 eV) generated by LED source, with 1 W power. By choosing the wavelength in the visible range (3.26 (Violet) to 1.59 eV (red)), we should be able to excite the transitions at 1.7 eV ($^4T_2$) and 2.3 eV ($^4T_1$) in CrCl$_3$ and the transitions at 2.0 and 2.7 eV in the case of CrI$_3$. The latter transitions in CrI$_3$ correspond to ligand-to-metal charge transfer excitations. These transitions are optically active as well as spin allowed. Hence, they are expected to have a strong magneto-optical response[7].

Continuous wave (cw) X-band (9.45 GHz) ESR experiments were performed with a Bruker ELEXSYS II E500 ESR spectrometer (Bruker Biospin, Rheinstetten, Germany) equipped with a TE102 rectangular ESR resonator (Bruker ER 4102ST). A helium gas-flow cryostat (ICE Oxford, UK) and an ITC503 from Oxford Instruments, UK, were used for measurements at cryogenic temperatures. Additionally, all ESR experimental settings were kept constant for reproducibility. For all ESR measurements, the samples were inserted into a quartz tube under nitrogen. All the samples were carefully handled with nonmagnetic capsules to avoid contamination. The light source was a Solis-3c white (daylight) LED (Thorlabs). The sample was first cooled down to 10 K in the dark and the temperature was successively increased in steps. Then it was cooled down again to check for reproducibility/sample stability in several steps. It should be noted that the samples showed no sign of degradation before and after photo-excitation. Finally, all quantum chemistry calculations were performed using the MOLPRO package[8].

## Additional ESR and Magnetic Measurements:

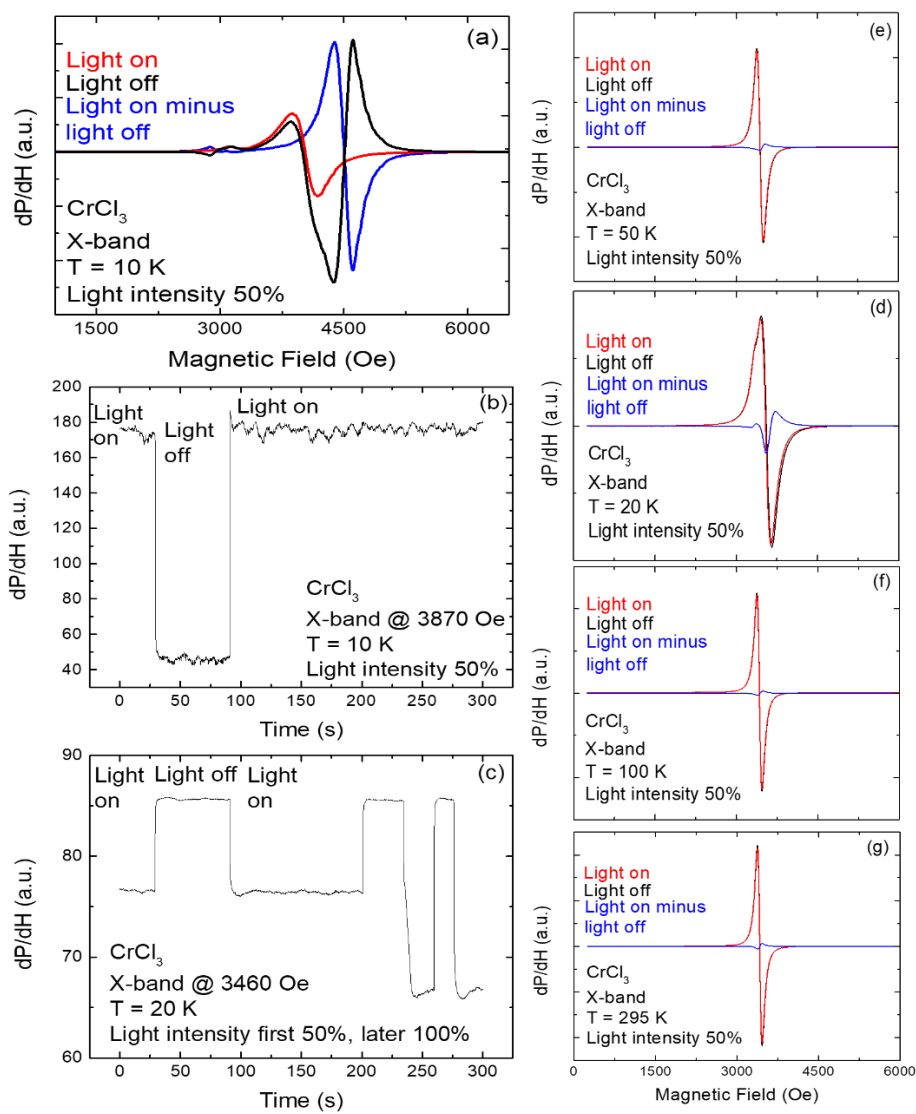

**Supplementary Figure 1:** Figure 1 (a) and (d) present the X-band (9.45 GHz) ESR spectra (0-7 kOe) recorded on $CrCl_3$ at 10 K and 20 K when the light is on (shown in red) and off (shown in black) in the ferromagnetic phase, respectively. Also included is the light on minus light off signal (shown in blue). The ESR signals in the paramagnetic phase recorded at 50 K, 100 K, and 295 K are shown in Figures 1(e) (f) (g), respectively. In Figures 1(b) and 1(c), we plotted the time traces of ESR signal intensity collected on $CrCl_3$ at 10 K and 20 K, at the resonance fields of 3870 Oe and 3460 Oe, respectively. An intensity of 100% corresponds to a fluence of ca. 40 mW/mm$^2$ at the sample.

In Supplementary Figure 2, we show the first derivative ESR spectra collected from $CrCl_3$ measured at 10 K before (light-off, shown in black) and after photo-excitation (light-on, shown in black). Also shown (in blue and red) are the fits generated by computer using the Dysonian line shape, shown below.

$$P = \left(\frac{16I}{\pi}\right)\left(\frac{a(\sqrt{2}H_{pp})^2 - 4a(H-H_o)^2 - 8(\sqrt{2}H_{pp})(H-H_o)}{(4(x-H_o)^2 + (\sqrt{2}H_{pp})^2)^2}\right) +$$
$$\left(\frac{16I}{\pi}\right)\left(\frac{a(\sqrt{2}H_{pp})^2 + 4a(H+H_o)^2 - 8(\sqrt{2}H_{pp})(H+H_o)}{(4(H+H_o)^2 + (\sqrt{2}H_{pp})^2)^2}\right) +$$

The parameters are as follows: $P$ is power absorbed, $I$ is a dimensionless intensity factor, $H_{PP}$ is a the peak-to-peak linewidth, $H_o$ is the resonance field, $c$ is a constant, and $H$ is the varying field[10–13]

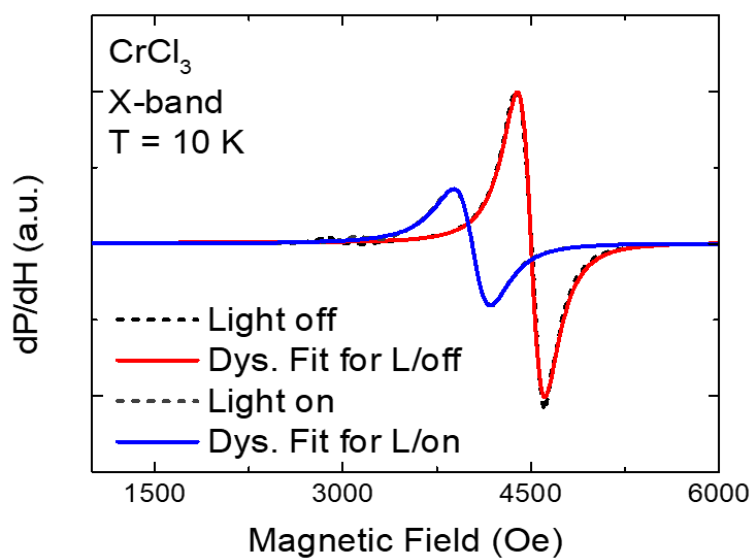

**Supplementary Figure 2.** Computer generated fits using the Dysonian lineshape applied to the experimental signal (in black) collected under dark (L/off, in red) and light (L/on, in blue)

In supplementary figure 3, ESR signals collected from $CrI_3$ were plotted at various temperatures 60 K (a), 65 K (b), and 100 K (c) with the light ON (red) and OFF (blue). The time traces of ESR signal intensity collected on $CrI_3$ at 60 K and 100 K (at the resonance field of 2650 Oe) are shown in Figs (d) and (e).

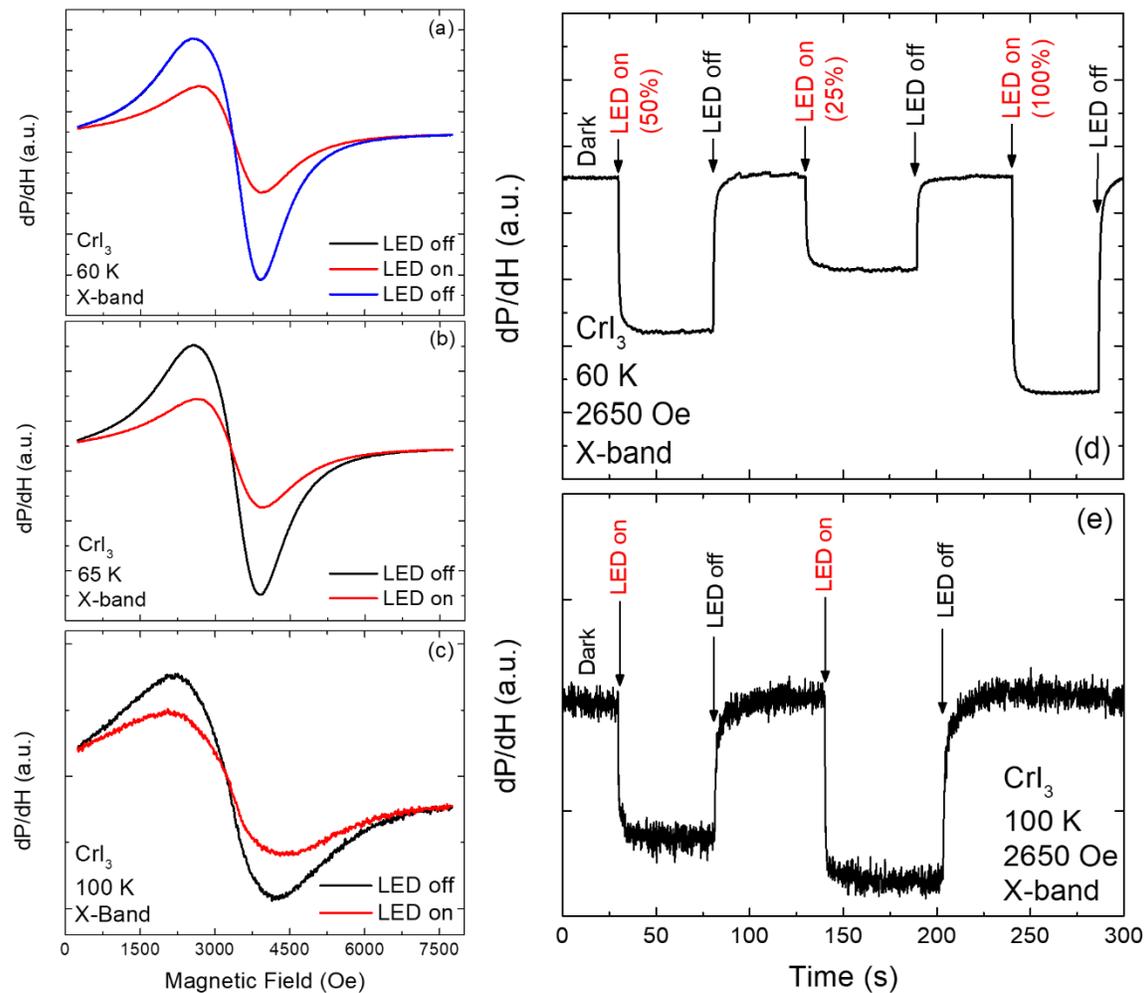

**Supplementary Figure 3.** Light-induced ESR signal of CrI$_3$ measured at 60K (a), 65K (b), 100K (c), and the time evolution of signal intensity measured at 60 K (d) and 100 K (e).

In Supplementary Figure 4, we show the first derivative ESR spectra collected from $CrI_3$ measured in the paramagnetic phase at 60 K (a), 65 K (b), and 100 K (c) before (light-off, shown in black) and after photo-excitation (light-on, shown in black). Also shown (in blue and red) are the fits generated by computer using the Dysonian line shape (see above).

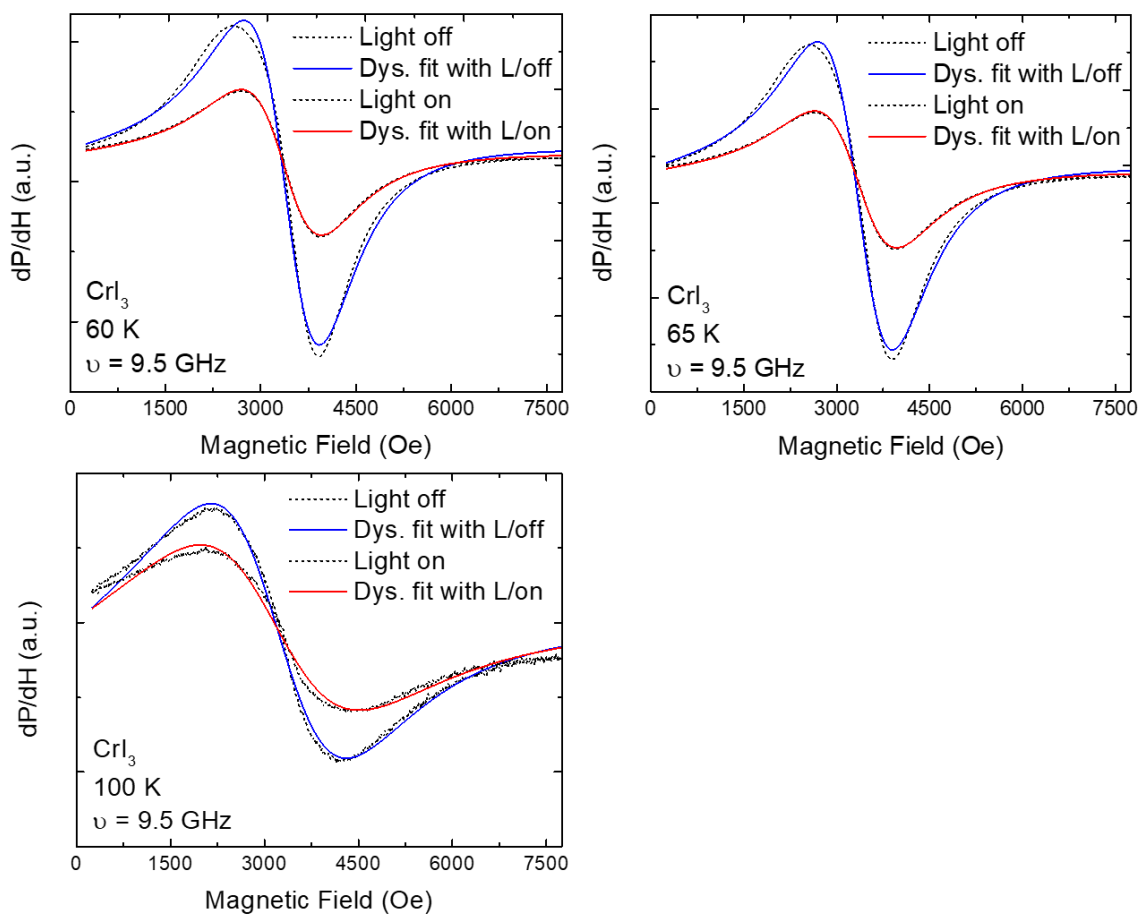

**Supplementary Figure 4**. Computer generated fits using the Dysonian lineshape applied to the experimental signal (in black) collected at 60 K (a), 65 K (b), and 100 K(c) (in the paramagnetic phase) under dark (L/off, in blue) and light (L/on, in blue)

In Supplementary Figure 5 (a,b), we have presented the ESR signals collected from CrI$_3$ in the ferromagnetic phase (<60 K) at two different temperatures 30 K (a) and 10 K (b). As one can immediately notice, the ESR signals are not reproduced upon photo-excitation. Most likely, it could be due to the strong interaction of light with magnetic domains which causes instability of ferromagnetic domains in the CrI$_3$ crystal.

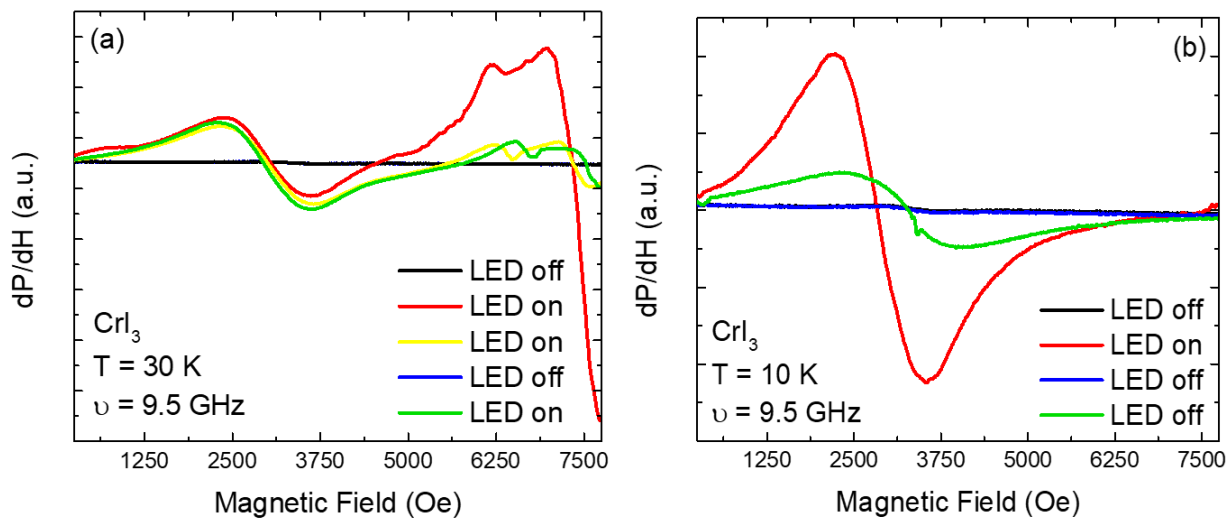

**Supplementary Figure 5.** ESR signals collected from CrI$_3$ at 30 K (a) and 10 K (b) in the ferromagnetic phase.

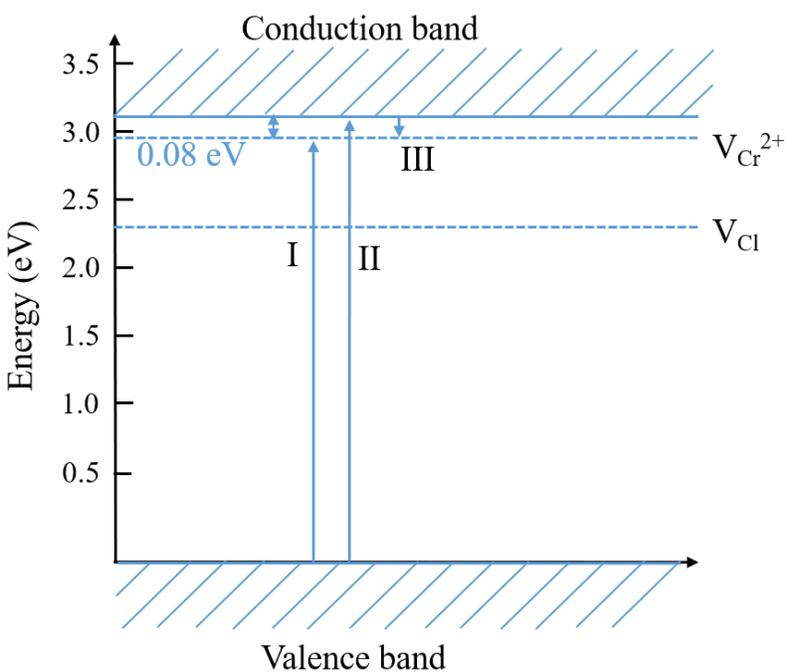

**Supplementary Figure 6.** Schematic for CrCl$_3$ demonstrating the band gap located at 3.1 eV along with the inter-band transition for Cr$^{2+}$ (3d$^4$) electrons. Transition I: Direct electronic transition from the valence band to Cr$^{2+}$ level; Transitions II+III are indirect transitions through conduction band.

## Quantum chemistry calculations:

*Ab initio* wavefunction based calculations were carried out on finite-size models using quantum chemistry package MOLPRO[14]. All-electron basis functions of quadruple-zeta quality were used for the Cr ions[15] in the reference unit while Cr$^{3+}$ sites belonging to the octahedra adjacent to the reference unit were described as closed-shell Cr$^{3+}$ t$^3_{2g}$ ions, using all-electron triple-zeta basis functions[15]. In the case of I ligands belonging to the reference unit, we used energy-consistent relativistic pseudopotentials along with triple-zeta quality basis sets for the valence shells[16], while for the I ligands belonging to the adjacent octahedra which are not shared with the central unit we relied on energy-consistent relativistic pseudopotentials and the double-zeta basis set[16]. In the case of Cl ligands belonging to the reference octahedron, we employed all-electron valence triple-zeta basis sets[17], while ligands of the adjacent octahedra that are not shared with the central octahedron were modeled with all-electron minimal atomic-natural-orbital basis sets[18].